\documentclass[aip,reprint]{revtex4-1}
\usepackage[T1]{fontenc}
\usepackage{times}
\usepackage{graphicx,color}
\usepackage{amsfonts,amsmath,amssymb,amsbsy}
\usepackage[colorlinks=true, linkcolor=blue, citecolor=blue, urlcolor=blue]{hyperref}

\let\mathbf=\boldsymbol

\def\emph#1{\textcolor{magenta}{#1}}
\begin{document}

\title{Current-driven skyrmionium in a frustrated magnetic system}

\author{Jing Xia}
\thanks{J. Xia and X. Zhang contributed equally to this work.}
\affiliation{School of Science and Engineering, The Chinese University of Hong Kong, Shenzhen, Guangdong 518172, China}
\affiliation{College of Physics and Electronic Engineering, Sichuan Normal University, Chengdu 610068, China}

\author{Xichao Zhang}
\thanks{J. Xia and X. Zhang contributed equally to this work.}
\affiliation{School of Science and Engineering, The Chinese University of Hong Kong, Shenzhen, Guangdong 518172, China}

\author{Motohiko Ezawa}
\email[Email:~]{ezawa@ap.t.u-tokyo.ac.jp}
\affiliation{Department of Applied Physics, The University of Tokyo, 7-3-1 Hongo, Tokyo 113-8656, Japan}

\author{Oleg A. Tretiakov}
\affiliation{School of Physics, The University of New South Wales, Sydney 2052, Australia}

\author{Zhipeng Hou}
\affiliation{South China Academy of Advanced Optoelectronics, South China Normal University, Guangzhou 510006, China}

\author{Wenhong Wang}
\affiliation{State Key Laboratory of Magnetism, Institute of Physics, Chinese Academy of Sciences, Beijing 100190, China}

\author{Guoping Zhao}
\affiliation{College of Physics and Electronic Engineering, Sichuan Normal University, Chengdu 610068, China}

\author{Xiaoxi Liu}
\affiliation{Department of Electrical and Computer Engineering, Shinshu University, 4-17-1 Wakasato, Nagano 380-8553, Japan}

\author{Hung T. Diep}
\affiliation{Laboratoire de Physique Th{\'e}orique et Mod{\'e}lisation, Universit{\'e} de Cergy-Pontoise, 95302 Cergy-Pontoise Cedex, France}

\author{Yan Zhou}
\email[Email:~]{zhouyan@cuhk.edu.cn}
\affiliation{School of Science and Engineering, The Chinese University of Hong Kong, Shenzhen, Guangdong 518172, China}

\begin{abstract}
Magnetic skyrmionium can be used as a nanometer-scale non-volatile information carrier, which shows no skyrmion Hall effect due to its special structure carrying zero topological charge. Here, we report the static and dynamic properties of an isolated nanoscale skyrmionium in a frustrated magnetic monolayer, where the skyrmionium is stabilized by competing interactions. The frustrated skyrmionium has a size of about $10$ nm, which can be further reduced by tuning perpendicular magnetic anisotropy or magnetic field. It is found that the nanoscale skyrmionium driven by the damping-like spin-orbit torque shows directional motion with a favored Bloch-type helicity. A small driving current or magnetic field can lead to the transformation of an unstable N{\'e}el-type skyrmionium to a metastable Bloch-type skyrmionium. A large driving current may result in the distortion and collapse of the Bloch-type skyrmionium. Our results are useful for the understanding of frustrated skyrmionium physics, which also provide guidelines for the design of spintronic devices based on topological spin textures.
\end{abstract}

\date{26 June 2020}


\maketitle


Topological spin textures have aroused significant interest within the field of magnetism and spintronics due to their highly possible spintronic applications~\cite{Bogdanov_1989,Roszler_NATURE2006,Nagaosa_NNANO2013,Wiesendanger_NATREVMAT2016,Finocchio_JPD2016,Kang_PIEEE2016,Kanazawa_AM2017,Wanjun_PHYSREP2017,Fert_NATREVMAT2017,Zhou_NSR2018,ES_JAP2018,Zhang_JPCM2020}, such as racetrack-type memories, logic computing gates, and neuromorphic computing devices.
Exemplary topological spin textures include skyrmions~\cite{Bogdanov_1989} and skyrmioniums~\cite{Bogdanov_JMMM1999,Finazzi_PRL2013}, which can be stabilized in magnetic materials with the Dzyaloshinskii-Moriya (DM) interaction~\cite{Muhlbauer_SCIENCE2009,Yu_NATURE2010,Woo_NMATER2016,Wanjun_NPHYS2017,Litzius_NPHYS2017}.
Recently, a number of studies have suggested that skyrmions can also be found in frustrated magnetic systems, where competing exchange interactions lead to the formation of skyrmions at certain conditions~\cite{Okubo_PRL2012,Leonov_NCOMMS2015,Lin_PRB2016A,Hayami_PRB2016A,Rozsa_PRL2016,Leonov_NCOMMS2017,Kharkov_PRL2017,Xichao_NCOMMS2017,Yuan_PRB2017,Hou_AM2017,Hu_SR2017,Malottki_SR2017,Liang_NJP2018,Ritzmann_NE2018,Kurumaji_SCIENCE2019,Desplat_PRB2019,Xia_PRApplied2019,Zarzuela_PRB2019,Lohani_PRX2019,Diep_Entropy2019,Diep_2020,Zhang_PRB2020,Diep_AIP2013,Diep_PRB2019,Hou_AM2020}.
The magnetic skyrmions stabilized by frustrated exchange interactions show same topological configurations but different dynamic properties to that stabilized by DM interactions~\cite{Leonov_NCOMMS2015,Lin_PRB2016A,Leonov_NCOMMS2017,Kharkov_PRL2017,Xichao_NCOMMS2017,Ritzmann_NE2018,Xia_PRApplied2019,Zhang_PRB2020}, such as circular motion and helicity rotation.
However, the physical properties of skyrmioniums in frustrated magnetic systems remain elusive.

A magnetic skyrmion is formed by a circular domain wall, while a magnetic skyrmionium is formed by two concentric circular domain walls~\cite{Bogdanov_JMMM1999,Finazzi_PRL2013,Rohart_PRB2013,Leonov_EPJ2014,Beg_SREP2015,Komineas_PRB2015,Komineas_2015,Liu_AIPAdv2015,LiuYan_PRB2015,Xichao_PRB2016,Zheng_PRL2017,Fujita_PRB2017,Kolesnikov_SciRep2018,Lisai_ApplPhysLett2018,Shen_ApplPhysLett2018,Zhang_NanoLett2018,Gobel_SciRep2019,Song_APE2019,Bo_JPDAP2020,Ishida_2020,Beg_2017,Loreto_2018,Foster_2019}.
A single isolated skyrmion carries a integer topological charge~\cite{Nagaosa_NNANO2013,Wiesendanger_NATREVMAT2016,Finocchio_JPD2016,Kang_PIEEE2016,Kanazawa_AM2017,Wanjun_PHYSREP2017,Fert_NATREVMAT2017,Zhou_NSR2018,ES_JAP2018,Zhang_JPCM2020}, which is defined as
$Q=\int\boldsymbol{m}(\boldsymbol{r})\cdot\left(\partial_{x}\boldsymbol{m}(\boldsymbol{r})\times\partial_{y}\boldsymbol{m}(\boldsymbol{r})\right)d^{2}\boldsymbol{r}/4\pi$.
However, the skyrmionium carries a topological charge of $Q=0$ and it can be seen as a combination of two skyrmions with opposite topological charges, i.e., $Q=+1$ and $Q=-1$.
The skyrmionium can be created by electric and optical methods~\cite{Finazzi_PRL2013,Xichao_PRB2016,Fujita_PRB2017,Kolesnikov_SciRep2018,Gobel_SciRep2019}.
Because of the absence of topological charge, a rigid isolated skyrmionium shows dynamics that is independent of its topological structure.
Hence, the most important feature is that a skyrmionium with $Q=0$ shows no skyrmion Hall effect~\cite{Zhang_PRB2016,Wanjun_NPHYS2017,Litzius_NPHYS2017,Tretiakov_PRL2016}, which has been regarded as a promising feature for reliable in-line motion in narrow nanotracks~\cite{Zhang_PRB2016,Guoqiang_NL2017}.
As a result, the skyrmionium can be used as a nanoscale non-volatile information carrier in spintronic applications~\cite{Xichao_PRB2016,Lisai_ApplPhysLett2018,Shen_ApplPhysLett2018,Gobel_SciRep2019}, such as the racetrack-type memories~\cite{Zhang_PRB2016,Tomasello_SREP2014,Guoqiang_NL2017}.


In this work, we numerically explore the skyrmionium stabilized by frustrated exchange interactions in a magnetic monolayer with perpendicular magnetic anisotropy (PMA), which are important for practical applications based on the electric manipulation of skyrmioniums.
Our simulations are based on the $J_{1}$-$J_{2}$-$J_{3}$ classical Heisenberg model on a simple monolayer square lattice~\cite{Lin_PRB2016A,Xichao_NCOMMS2017,Xia_PRApplied2019,Zhang_PRB2020,Kaul_2004}, where three competing ferromagnetic (FM) and antiferromagnetic (AFM) Heisenberg exchange interactions lead to exchange frustration~\cite{Diep_Entropy2019}.
The simulations are performed by using the object oriented micromagnetic framework (OOMMF)~\cite{OOMMF} package upgraded with homemade extension modules for the $J_{1}$-$J_{2}$-$J_{3}$ exchange interactions~\cite{Lin_PRB2016A,Xichao_NCOMMS2017,Xia_PRApplied2019}.
The Hamiltonian $\mathcal{H}$ includes the the nearest-neighbor exchange ($J_1$), next-nearest-neighbor exchange ($J_2$), next-next-nearest-neighbor exchange ($J_3$), PMA ($K$), applied magnetic field ($\boldsymbol{B}$), and dipole-dipole interaction (DDI) terms (see supplementary material).

The spin dynamics is governed by the Landau-Lifshitz-Gilbert (LLG) equation augmented with the damping-like spin-orbit torque~\cite{OOMMF,Lin_PRB2016A,Xichao_NCOMMS2017,Xia_PRApplied2019,Zhang_PRB2020,Tomasello_SREP2014}, given as
\begin{equation}
\frac{d\boldsymbol{m}}{dt}=-\gamma_{0}\boldsymbol{m}\times\boldsymbol{h}_{\rm{eff}}+\alpha\left(\boldsymbol{m}\times\frac{d\boldsymbol{m}}{dt}\right)-u\boldsymbol{m}\times(\boldsymbol{m}\times\boldsymbol{p}),
\label{eq:LLG}
\end{equation}
where $|\boldsymbol{m}|=1$ represents the normalized spin,
$\boldsymbol{h}_{\rm{eff}}=-\frac{1}{\mu_{0}M_{\text{S}}}\cdot\frac{\delta\mathcal{H}}{\delta\boldsymbol{m}}$ is the effective field,
$t$ is the time,
$\alpha$ is the Gilbert damping parameter,
and $\gamma_0$ is the absolute gyromagnetic ratio.
$u=\left|\left(\gamma_{0}\hbar/\mu_{0}e\right)\right|\cdot\left(j\theta_{\text{SH}}/2a M_{\text{S}}\right)$ is the spin torque coefficient,
where $\hbar$ is the reduced Planck constant, $e$ is the electron charge, $\mu_{0}$ is the vacuum permeability constant, $a$ is the lattice constant,
$j$ is the applied current density, $\theta_{\text{SH}}$ is the spin Hall angle, and $M_{\text{S}}$ is the saturation magnetization.
$\boldsymbol{p}=+\hat{y}$ denotes the spin current polarization direction.

The default geometry of the monolayer consists of $51 \times 51$ spins, and the lattice constant is $a=0.4$ nm (i.e., the mesh size is $0.4 \times 0.4 \times 0.4$ nm$^3$).
The default simulation parameters are given as~\cite{Xichao_NCOMMS2017,Xia_PRApplied2019,Zhang_PRB2020}:
$J_1=30$ meV,
$J_2=-0.8$ (in units of $J_1=1$),
$J_3=-0.9$ (in units of $J_1=1$),
$K=0.01$ (in units of $J_{1}/a^{3}=1$),
$B=0$ (in units of $J_{1}/a^{3}M_{\text{S}}=1$),
$\theta_{\text{SH}}=0.2$,
$\alpha=0.3$,
$\gamma_0=2.211\times 10^{5}$ m A$^{-1}$ s$^{-1}$,
and $M_{\text{S}}=580$ kA m$^{-1}$.
We have simulated the metastability diagram, which shows the frustrated skyrmionium can be a metastable state for a wide range of $J_2$, $J_3$, $K$, and $M_{\text{S}}$ parameters (see supplementary material).
Note that the minimum required value of $J_3$ for stabilizing skyrmioniums decreases with increasing magnitude of $J_2$ since both $J_2$ and $J_3$ are AFM exchange interactions that compete with FM $J_1$.

\begin{figure}[t]
\centerline{\includegraphics[width=0.45\textwidth]{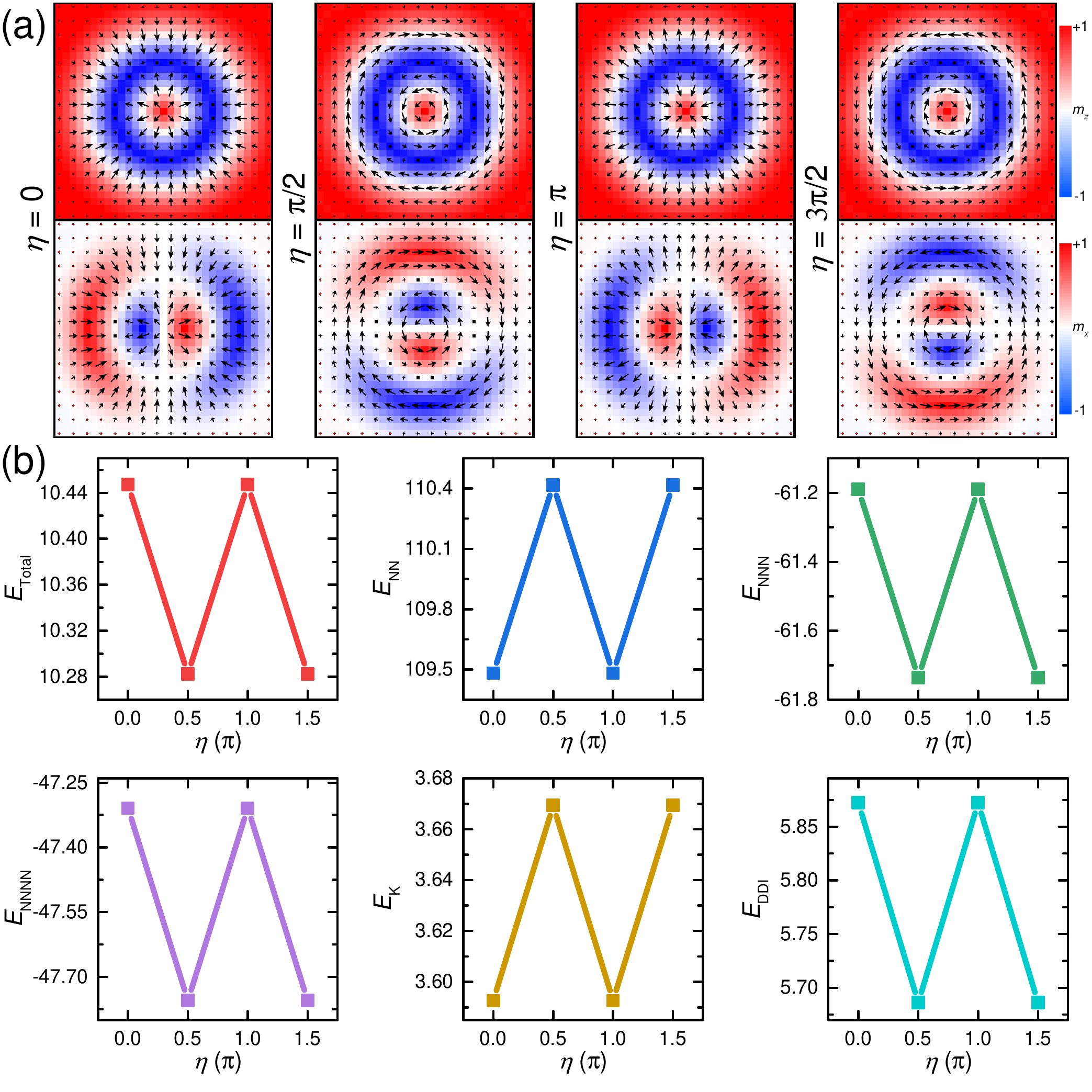}}
\caption{%
(a) Zoomed top view of relaxed skyrmioniums with different helicity $\eta$ in the frustrated magnetic monolayer. Here, $J_2=-0.8$, $J_3=-0.9$, $K=0.01$, and $B=0$. The arrows represent the spin directions. A single displayed arrow is a sampling of two real spins. The out-of-plane ($m_z$) or in-plane ($m_x$) spin component is color coded.
The size of the zoomed area is $12.4\times 12.4$ nm$^{2}$.  
(b) Energies of relaxed skyrmioniums with $Q=0$ and different $\eta$ in the frustrated magnetic monolayer with PMA ($K=0.01$).
The energies are given in units of $J_1=1$.
}
\label{FIG1}
\end{figure}

\begin{figure}[t]
\centerline{\includegraphics[width=0.45\textwidth]{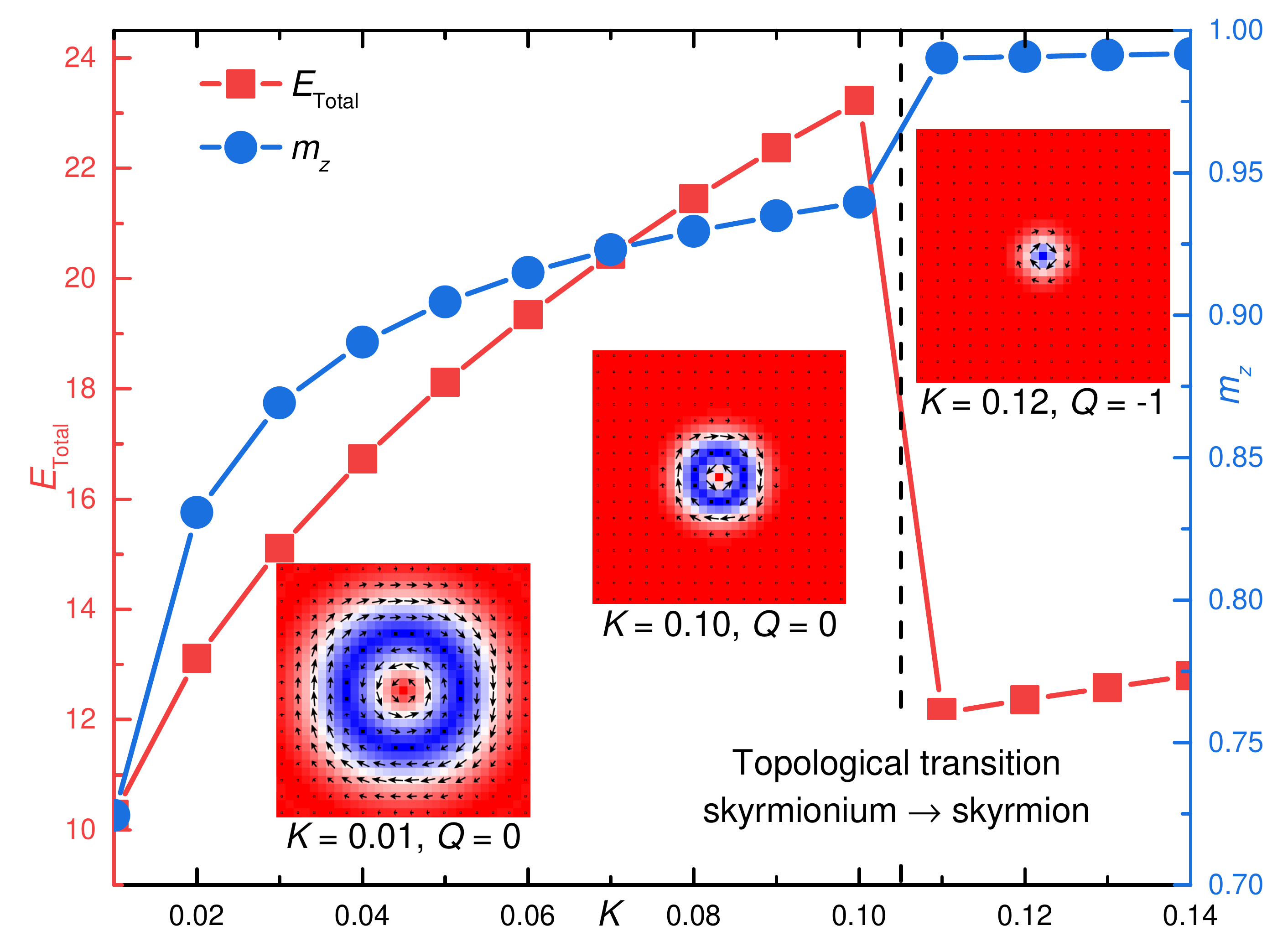}}
\caption{%
Total energy $E_{\text{Total}}$ and out-of-plane spin component $m_{z}$ as functions of PMA $K$. The initial state is a relaxed Bloch-type skyrmionium with $\eta=\pi/2$ at the monolayer center. Here, $J_2=-0.8$, $J_3=-0.9$, and $B=0$.
The energies are given in units of $J_1=1$.
Insets show zoomed top views ($12.4\times 12.4$ nm$^{2}$) of relaxed states at selected values of $K$.
}
\label{FIG2}
\end{figure}


As shown in Fig.~\ref{FIG1}, we first study the static properties of a relaxed isolated skyrmionium in the frustrated magnetic monolayer with PMA of $K=0.01$, where we set $J_2=-0.8$, $J_3=-0.9$, and $B=0$.
The spin texture is parametrized by
$\boldsymbol{m}(\boldsymbol{r})=\boldsymbol{m}(\theta,\phi)=(\sin\theta\cos\phi,\sin\theta\sin\phi,\cos\theta)$
with
$\phi=Q_{\text{v}}\psi+\eta$,
where $\psi$ is the azimuthal angle in the $x$-$y$ plane ($0\le\psi<2\pi$).
For the skyrmionium, we assume that $\theta$ rotates $2\pi$ for spins from the skyrmionium center to skyrmionium edge~\cite{Bogdanov_JMMM1999,Zhang_JPCM2020}.
It is worth mentioning that $\theta$ only rotates $\pi$ for the case of skyrmion~\cite{Bogdanov_1989,Bogdanov_JMMM1999,Zhang_JPCM2020}.
Hence, $Q_{\text{v}}=\frac{1}{2\pi}\oint_{C}d\phi$ is the vorticity and $\eta$ is the helicity defined mod $2\pi$, which describe the out-of-plane and in-plane structures of a skyrmionium, respectively. Note that the state with $\eta=0$ is identical to that with $\eta=2\pi$.

Fig.~\ref{FIG1}(a) shows spin configurations of relaxed skyrmioniums with $Q=0$ but different helicity $\eta=0,\pi/2,\pi,3\pi/2$. The skyrmionium has two concentric circular domain walls. The in-plane spin configuration of the inner circular domain wall is directly described by the helicity $\eta$, while the helicity of the outer circular domain wall is always equal to $\eta+\pi$. Once the helicity $\eta$ is given, the in-plane spin configurations of both inner and outer circular domain walls are determined.

The energies of relaxed skyrmioniums with $\eta=0,\pi/2,\pi,3\pi/2$ are given in Fig.~\ref{FIG1}(b). The total energy and all constituting energy terms depend on the helicity. The Bloch-type skyrmioniums with $\eta=\pi/2,3\pi/2$ are more energetically stable than the N{\'e}el-type skyrmioniums with $\eta=0,\pi$.
The total energy difference is mainly induced by the DDI, which naturally favors Bloch-type spin configurations. The skyrmionium energy will be independent of $\eta$ if the effect of DDI is not taken into account (see supplementary material).

We also study the effect of PMA on the static profile of a relaxed skyrmionium.
Figure~\ref{FIG2} shows the spin configuration of a Bloch-type skyrmionium with $\eta=\pi/2$, which is relaxed in the frustrated magnetic monolayer for different $K$.
The size of skyrmionium decreases with increasing $K$.
The diameter of the outer circular domain wall of the skyrmionium is $\sim 9.2$ nm at $K=0.01$, while it is only $\sim 4.4$ nm at $K=0.10$.
In particular, when $K\geq 0.11$, the skyrmionium with $Q=0$ is transformed to a skyrmion with $Q=-1$, which is a result of the annihilation of the inner circular domain wall. A further increase in $K$ will lead to the annihilation of the skyrmion.
The total energy of the skyrmionium also increases with $K$. When the skyrmionium is transformed to a skyrmion, the total energy suddenly drops due to the collapse of the inner circular domain wall.
Besides, the skyrmionium size can be adjusted by a perpendicular magnetic field $B$ within certain range (see supplementary material).

\begin{figure}[t]
\centerline{\includegraphics[width=0.45\textwidth]{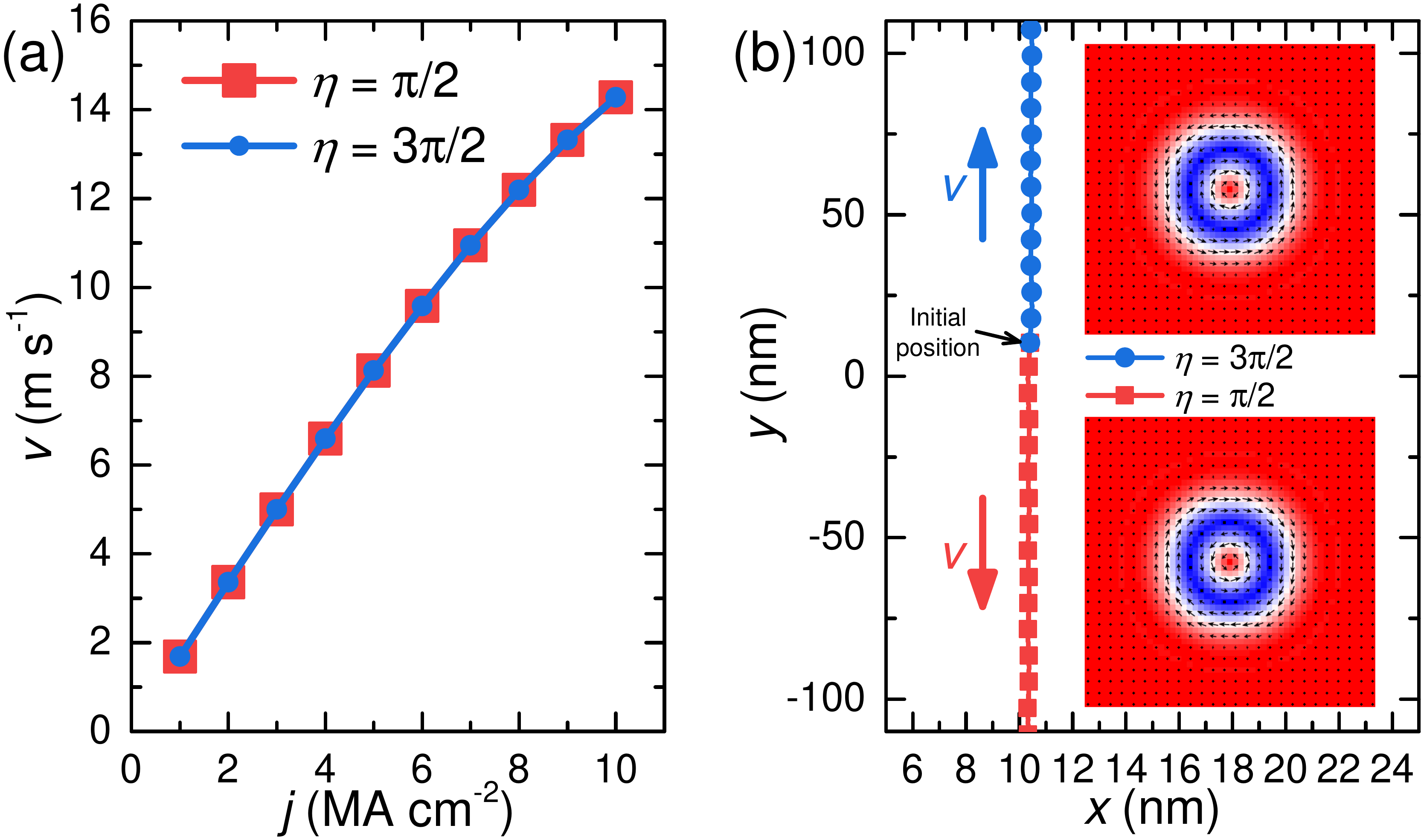}}
\caption{%
(a) Velocities $v$ for isolated Bloch-type skyrmioniums with $\eta=\pi/2,3\pi/2$ driven by the damping-like spin-orbit torque as functions of the driving current density $j$. The velocity at different $j$ is measured when steady motion is attained. Here, $J_2=-0.8$, $J_3=-0.9$, $K=0.01$, and $B=0$.
(b) Trajectories for isolated Bloch-type skyrmioniums with $\eta=\pi/2,3\pi/2$ driven by the damping-like spin-orbit torque. Here, $j=5$ MA cm$^{-2}$. The red and blue arrows indicate the direction of motion. Inset shows top views ($20.4\times 20.4$ nm$^{2}$) of current-driven skyrmioniums with $\eta=\pi/2,3\pi/2$.
(Multimedia view)
}
\label{FIG3}
\end{figure}

We continue to investigate the dynamics of metastable Bloch-type skyrmioniums driven by the damping-like spin-orbit torque (see Eq.~\ref{eq:LLG}).
We assume that the damping-like spin-orbit torque is generated by utilizing the spin Hall effect of a heavy-metal substrate~\cite{Tomasello_SREP2014,Kang_PIEEE2016,Wanjun_PHYSREP2017,Fert_NATREVMAT2017,ES_JAP2018,Zhang_JPCM2020}.
For the sake of simplicity, we ignore the effect of the field-like torque, as it cannot drive skyrmions and skyrmioniums into directional motion~\cite{Litzius_NPHYS2017}.

Figure~\ref{FIG3}(a) shows the velocities of Bloch-type skyrmioniums with $\eta=\pi/2,3\pi/2$ driven by the damping-like spin-orbit torque.
The Bloch-type skyrmionium with $\eta=\pi/2$ moves toward the $-y$ direction, while the one with $\eta=3\pi/2$ moves toward the $+y$ direction [see Fig.~\ref{FIG3}(b) and multimedia view]. Namely, the direction of motion is parallel to the spin polarization direction $\boldsymbol{p}=+\hat{y}$, and depends on the helicity of skyrmionium.
However, the velocity is independent of helicity. The Bloch-type skyrmioniums with $\eta=\pi/2$ and $\eta=3\pi/2$ show identical current-velocity relation~\cite{Muller_2020}, where the velocity is proportional to the driving current density $j$.
It is worth mentioning that the Bloch-type skyrmioniums moves at speed of $\sim 14.3$ m s$^{-1}$ at $j=10$ MA cm$^{-2}$. However, as reported in Ref.~\onlinecite{Xichao_PRB2016}, the skyrmioniums stabilized by DM interactions in conventional FM systems could reach a speed of $\sim 92.2$ m s$^{-1}$ at $j=10$ MA cm$^{-2}$, which is faster than the frustrated skyrmioniums. The reason is that the speed induced by the same current density increases with the size of skyrmionium~\cite{Xichao_PRB2016}. In this work, the diameter of the outer circular domain wall of the frustrated skyrmionium at $j=10$ MA cm$^{-2}$ is $\sim 9$ nm, while it is $\sim 90$ nm in Ref.~\onlinecite{Xichao_PRB2016}.
The larger size of skyrmionium leads to higher speed, however, it may result in lower storage density of skyrmionium-based memory applications.
On the other hand, the skyrmionium in the frustrated magnetic monolayer driven by a moderate damping-like torque (i.e., $j=1\sim 10$ MA cm$^{-2}$) shows directional motion instead of circular motion, which is different from frustrated skyrmions driven by the damping-like spin-orbit torque~\cite{Xichao_NCOMMS2017,Xia_PRApplied2019}.

\begin{figure}[t]
\centerline{\includegraphics[width=0.45\textwidth]{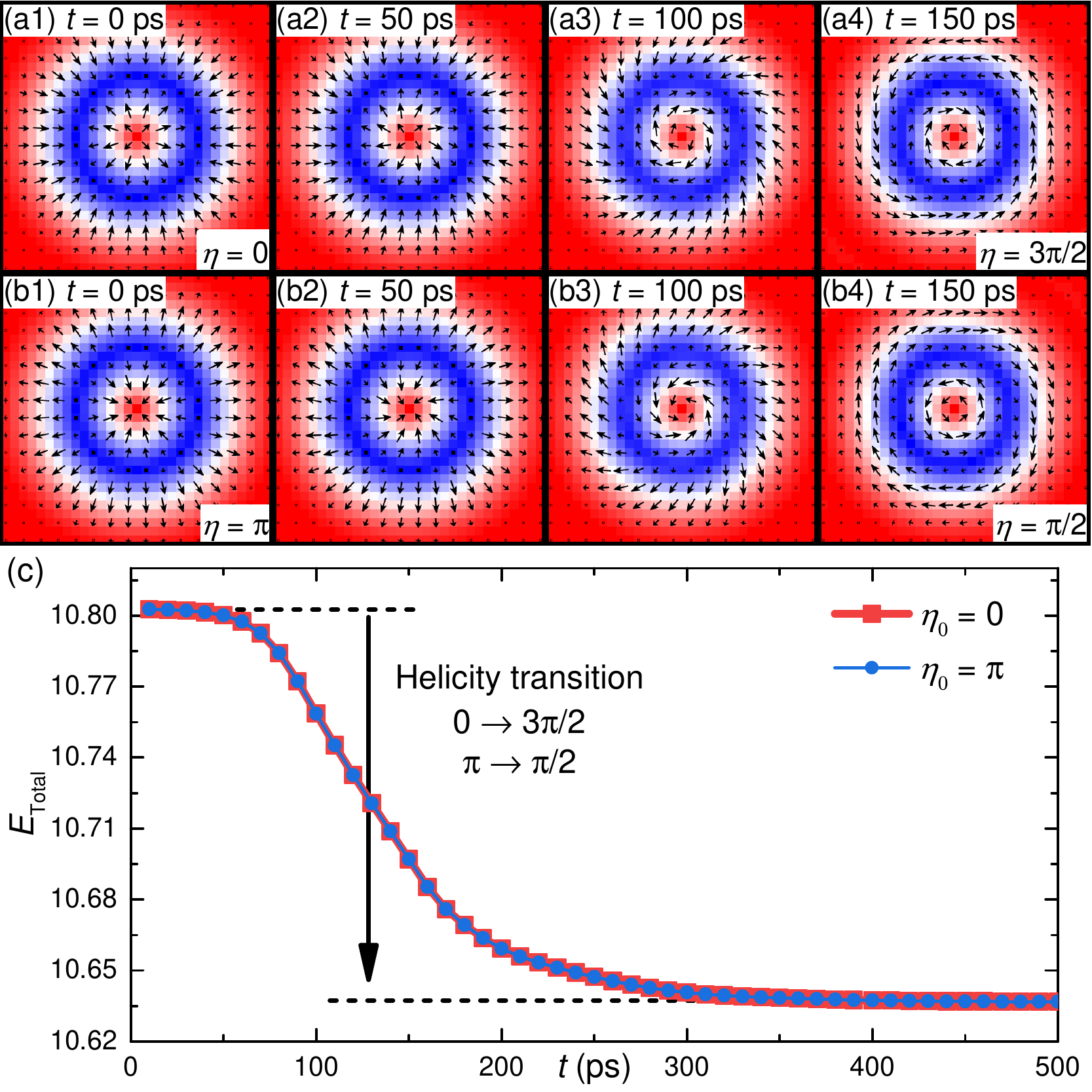}}
\caption{%
Small-current-induced transition from unstable N{\'e}el-type skyrmioniums to metastable Bloch-type skyrmioniums.
[(a)-(b)] Zoomed top view ($12.4\times 12.4$ nm$^{2}$) of the skyrmionium at selected times. Here, $j=1$ MA cm$^{-2}$, $J_2=-0.8$, $J_3=-0.9$, $K=0.01$, and $B=0$. The initial helicity of the skyrmionium is $\eta_{0}=0$ or $\eta_{0}=\pi$. The arrows represent the spin directions. A single displayed arrow is a sampling of two real spins.
(c) Total energy $E_{\text{Total}}$ as a function of time $t$.
The energies are given in units of $J_1=1$.
(Multimedia view)
}
\label{FIG4}
\end{figure}

If the initial state is a relaxed but unstable N{\'e}el-type skyrmionium with $\eta=0$ or $\eta=\pi$, a small value of damping-like spin-orbit torque acting on the skyrmionium will lead to the transition of the unstable N{\'e}el-type helicity to metastable Bloch-type helicity. As shown in Fig.~\ref{FIG4}(a), when a small driving current of $j=1$ MA cm$^{-2}$ is applied, the skyrmionium with $\eta=0$ is transformed to a skyrmionium with $\eta=3\pi/2$ (see multimedia view). Also, the skyrmionium with $\eta=\pi$ is transformed to a skyrmionium with $\eta=\pi/2$ [see Fig.~\ref{FIG4}(b)].
The total energy of the system decreases during the current-induced helicity transition [see Fig.~\ref{FIG4}(c)] as Bloch-type skyrmioniums are more energetically favorable.
Indeed, there is no energy barrier between the Bloch-type and N{\'e}el-type frustrated skyrmioniums (see supplementary material), and such a transition process can also be induced by a magnetic field pulse (see supplementary material) or other external stimuli, such as the Oersted field~\cite{Yin_2016}.

\begin{figure}[t]
\centerline{\includegraphics[width=0.45\textwidth]{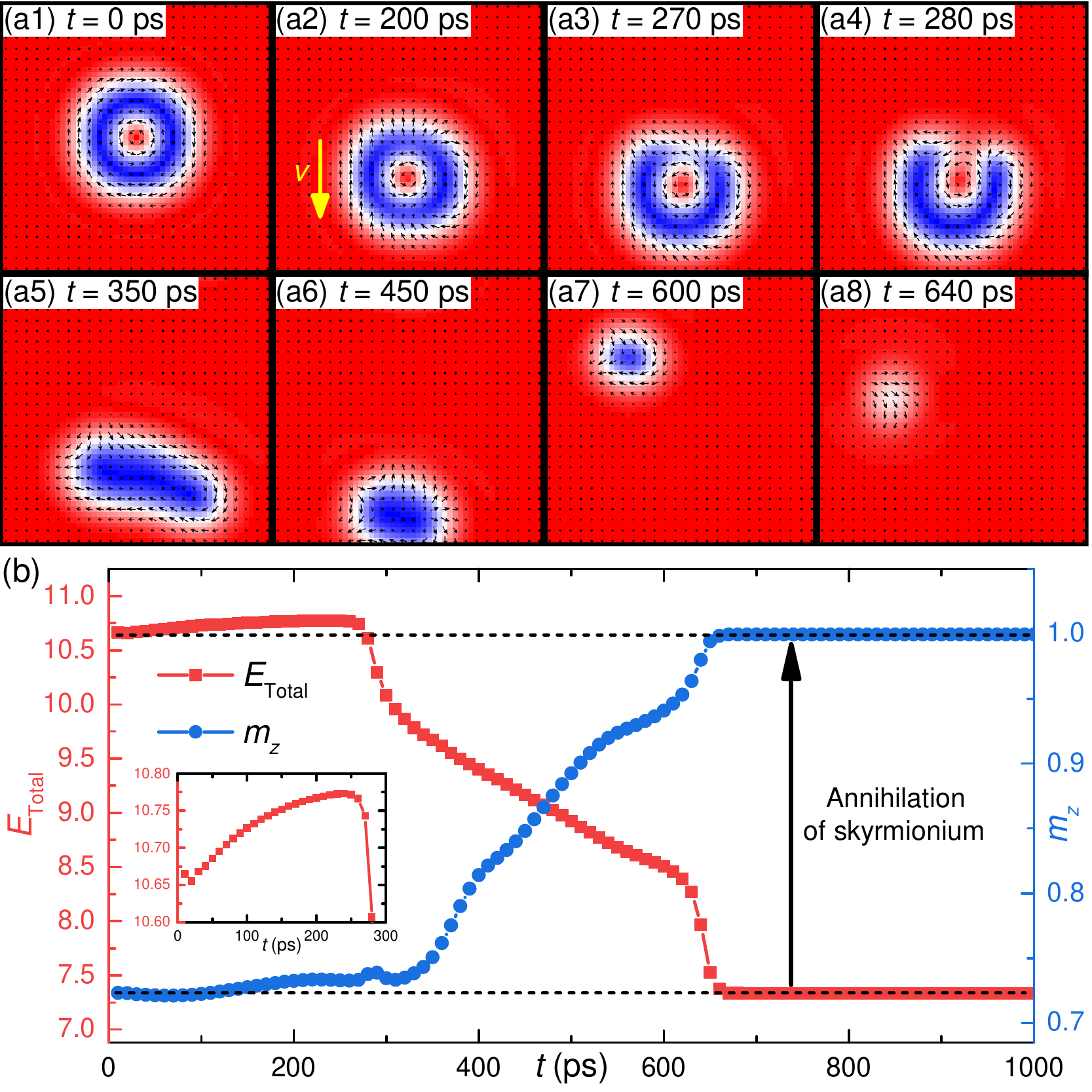}}
\caption{%
Large-current-induced distortion and annihilation of metastable Bloch-type skyrmioniums.
(a) Top view ($20.4\times 20.4$ nm$^{2}$) of the skyrmionium at selected times. Here, $j=20$ MA cm$^{-2}$, $J_2=-0.8$, $J_3=-0.9$, $K=0.01$, and $B=0$. The initial helicity of the skyrmionium is $\eta_{0}=\pi/2$. The arrows represent the spin directions. A single displayed arrow is a sampling of two real spins. The yellow arrow denotes the direction of motion.
(b) Total energy $E_{\text{Total}}$ and out-of-plane spin component $m_{z}$ as functions of time $t$. Inset shows zoomed view of $E_{\text{Total}}$ for $t=0\sim 300$ ps.
The energies are given in units of $J_1=1$.
(Multimedia view)
}
\label{FIG5}
\end{figure}

In Fig.~\ref{FIG5}, we demonstrate that a large driving current can result in the distortion and annihilation of metastable Bloch-type skyrmioniums during their motion (see multimedia view).
Figure~\ref{FIG5}(a) shows that when a large driving current of $j=20$ MA cm$^{-2}$ is applied, the metastable Bloch-type skyrmionium with $\eta=\pi/2$ moves toward the $-y$ direction. However, the spin configuration of its outer circular domain wall is distorted soon after the injection of the driving current.
At $t=270$ ps, the outer and inner circular domain walls touch each other and the skyrmionium structure is therefore transformed into a topological trivial bubble with $Q=0$.
The topological trivial bubble with $Q=0$ continues to move and shrink, which is ultimately annihilated at $t=640$ ps. The final state of the system is the FM state.
During the large-current-induced distortion and annihilation of the Bloch-type skyrmionium, the total energy of the system and the out-of-plane spin component decreases and increases with time, respectively [see Fig.~\ref{FIG5}(b)].
It is noteworthy that the total energy of the system slightly increases before the destruction of the skyrmionium structure [see Fig.~\ref{FIG5}(b) inset], which indicates the metastable Bloch-type skyrmionium is protected by an energy barrier.

In conclusion, we have studied the static and dynamic properties of an isolated skyrmionium in a frustrated magnetic monolayer.
Possible materials for hosting frustrated skyrmioniums are dihalides Fe$_x$Ni$_{1-x}$Br$_2$, which can have an easy-axis anisotropy along the $c$ axis~\cite{Regnault_1982,Moore_1985,Leonov_NCOMMS2015}. The low-dimensional compound Pb$_2$VO(PO$_4$)$_2$ is also a candidate material due to its frustrated square lattice with small FM nearest-neighbor and large AFM next-nearest-neighbor exchange interactions~\cite{Kaul_2004}.
In conventional FM and AFM materials, the stabilization of skyrmionium is usually realized in the presence of the interfacial or bulk DM interaction, while the DM interaction is not required to stabilize the frustrated skyrmionium.

We find that the skyrmionium energy depends on its helicity. Namely, the Bloch-type skyrmioniums with $\eta=\pi/2,3\pi/2$ are metastable, while the N{\'e}el-type skyrmioniums are unstable.
The size of a frustrated skyrmionium could be smaller than $10$ nm, which is much smaller than the typical sizes of skyrmions and skyrmioniums in conventional FM materials (i.e., $50\sim 500$ nm). The nanoscale size of frustrated skyrmionium is an advantage for increasing information storage density of skyrmionium-based memories.
We also find that the Bloch-type skyrmioniums can be driven into steady linear motion by a moderate damping-like spin-orbit torque, where the direction of motion depends on the skyrmionium helicity and the velocity increases with the current density.
The directional motion of skyrmionium can be utilized to build a racetrack-type device, and the helicity-dependent dynamics provides a possibility for realizing combined memory and computing functions, which is worth further investigation.

We also demonstrate that a small current or magnetic field can lead to the transformation of an unstable N{\'e}el-type skyrmionium to a metastable Bloch-type skyrmionium, and a large current can result in the annihilation of the Bloch-type skyrmionium during its motion.
The current-induced annihilation can be used as an effective method for simultaneously erasing all skyrmionium bits in racetrack-type memory applications. Note that in the skyrmion-based racetrack-type memory, the erasing of skyrmion bits is realized in a one-by-one manner by driving all skyrmions out of the racetrack end.
We believe our results are useful for understanding the frustrated skyrmionium physics, and can provide guidelines for the design of spintronic devices based on skyrmioniums.


See supplementary material for the metastability diagram and more simulation results. The data that support the findings of this study are available from the corresponding authors upon reasonable request.


\vbox{}

This work was supported by
the President's Fund of CUHKSZ,
the Longgang Key Laboratory of Applied Spintronics, 
the Shenzhen Peacock Group Plan (Grant No. KQTD20180413181702403),
the Guangdong Basic and Applied Basic Research Foundation (Grant No. 2019A1515110713),
the National Natural Science Foundation of China (Grant Nos. 11974298, 61961136006, 51901081, 11604148, 11874410, 51771127, 51571126, and 51772004),
the National Key R\&D Program of China (Grant Nos. 2017YFA0303202 and 2017YFA206303),
the Key Research Program of the Chinese Academy of Sciences (Grant No. KJZD-SW-M01),
the Grants-in-Aid for Scientific Research from JSPS KAKENHI (Grant Nos. JP18H03676, JP17K05490, 17K19074, 26600041, and 22360122),
the CREST of JST (Grant Nos. JPMJCR16F1 and JPMJCR1874),
the Australian Research Council (Grant No. DP200101027),
and the Cooperative Research Project Program at the Research Institute of Electrical Communication, Tohoku University.

\end{document}